\documentclass[10pt]{article}
\input epsf

\newcommand{\news}{\setcounter{equation}{0}}
\newcommand{\be}{\begin{equation}}
\newcommand{\ee}{\end{equation}}
\newcommand{\bea}{\begin{eqnarray}}
\newcommand{\eea}{\end{eqnarray}}
\newcommand{\bean}{\begin{eqnarray*}}
\newcommand{\eean}{\end{eqnarray*}}
\font\upright=cmu10 scaled\magstep1
\font\sans=cmss12
\newcommand{\ssf}{\sans}
\newcommand{\stroke}{\vrule height8pt width0.4pt depth-0.1pt}
\newcommand{\Z}{\hbox{\upright\rlap{\ssf Z}\kern 2.7pt {\ssf Z}}}

\newcommand{\C}{{\rlap{\rlap{C}\kern 3.8pt\stroke}\phantom{C}}}
\newcommand{\R}{\hbox{\upright\rlap{I}\kern 1.7pt R}}
\newcommand{\CP}{\C{\upright\rlap{I}\kern 1.5pt P}}
\newcommand{\identity}{{\upright\rlap{1}\kern 2.0pt 1}}
\newcommand{\half}{\frac{1}{2}}
\newcommand{\pr}{\partial}

\newcommand{\M}{{\cal M}}
\newcommand{\e}{\epsilon}
\begin{document}
\pagestyle{plain}

\title{\vskip -70pt
\begin{flushright}
{\normalsize DAMTP-2008-32} \\
\end{flushright}
\vskip 50pt
{\bf \Large \bf One-Vortex Moduli Space and Ricci Flow}
 \vskip 30pt
}
\author{Nicholas S. Manton\thanks{email N.S.Manton@damtp.cam.ac.uk} \\[15pt]
{\normalsize
{\sl Department of Applied Mathematics and Theoretical Physics,}}\\
{\normalsize {\sl University of Cambridge,}}\\
{\normalsize {\sl Wilberforce Road, Cambridge CB3 0WA, England.}}\\
}
\vskip 20pt
\date{April 2008}
\maketitle
\vskip 20pt

\begin{abstract}
The metric on the moduli space of one abelian Higgs vortex on a
surface has a natural geometrical evolution as the Bradlow 
parameter, which determines the vortex size, varies. It is shown 
by various arguments, and by calculations in special cases, that 
this geometrical flow has many similarities to Ricci flow.
\end{abstract}

\vskip 80pt

\newpage
\section{Introduction}
\news

In the abelian Higgs model at critical coupling, defined in the plane,
there are static $N$-vortex solutions in which the Higgs field
vanishes at precisely $N$ (not necessarily distinct) points. The moduli 
space of solutions $\M_N$ is a manifold of complex dimension $N$. There is a 
natural K\"ahler metric on $\M_N$, and motion along a geodesic in the moduli 
space corresponds to an $N$-vortex motion at slow speeds,
approximating a solution of the time-dependent field equations 
\cite{Man1,Stu}. 

The abelian Higgs model can be straightforwardly extended to any smooth
surface $\Sigma$ without boundary, for example the hyperbolic plane, or
a compact Riemann surface. A metric on $\Sigma$ must be specified.
Provided the area of $\Sigma$ is sufficiently large, there is again 
a moduli space of $N$-vortex solutions, with a natural 
K\"ahler metric. 

In this paper we shall investigate the case of one vortex on a compact
surface $\Sigma$ with arbitrary metric. Little attention has
previously been paid to this apparently simple case. It is helpful here 
to use Bradlow's formulation of the vortex equations. Bradlow introduced 
an explicit positive parameter $\tau$ such that the area of a 
vortex is proportional to $\tau$ \cite{Brad}\footnote{In our convention, $\tau$
is the inverse of the parameter that appears in \cite{Brad}.}. In some of
the literature, $\tau$ is scaled to unity and the area $A$ of $\Sigma$ 
regarded as variable, with the ratio of $A$ to the area of a vortex the 
physically interesting quantity. However we shall not do this here. 
We shall consider the area $A$ as fixed, but $\tau$ as variable. 
1-vortex solutions exist provided $0 < \tau < \frac{A}{4\pi}$.  
The 1-vortex moduli space $\M_1$ we shall simply denote by $\M$ 
from now on. As a manifold, $\M$ is a copy of $\Sigma$, since given any point 
of $\Sigma$, there is a unique vortex solution whose Higgs field
vanishes at that point. The metric on $\M$ depends on $\tau$ and 
generally differs from the metric on $\Sigma$. For example, 
for a vortex on a round sphere, $\M$ is also a round sphere, 
by symmetry, but its area is smaller. For a vortex on a flat torus, 
the moduli space is the same flat torus.

Our intuition is that in the limit $\tau \to 0$, the metric on moduli
space will be the original metric on $\Sigma$, since a vortex is
pointlike in this limit, and should move along geodesics of $\Sigma$. 
Then, as $\tau$ increases, the moduli space 
metric will partly smooth out the wrinkles in the metric of $\Sigma$. 
This is because the vortex occupies a finite region of $\Sigma$. 
Its motion, according to the geodesic approximation, is along a
geodesic of $\M$, so the metric on $\M$ should depend 
on the metric on $\Sigma$ averaged in some way over the region of the vortex.
We shall study how the metric on $\M$ varies with $\tau$, and think 
of this as a geometrical flow, with $\tau$ regarded as an analogue of time.

Calculating the metric on moduli space is not possible explicitly, 
in general, but there are a number of mathematical results which 
determine some of its properties. Samols found a formula for the 
metric on $\M_N$ in terms of coefficients in the expansion of the 
Higgs field around each vortex centre \cite{Sam}. From this 
formula it follows 
that the metric is K\"ahler. For vortices on a compact Riemann 
surface $\Sigma$, the real cohomology class of the K\"ahler 2-form 
on $\M_N$ is known, and from this the volume of $\M_N$ can be
computed \cite{ManNas}. It depends on the area of $\Sigma$, the 
genus of $\Sigma$, and on $\tau$ and $N$. These results simplify 
in the 1-vortex case.

In this paper, we give the area of the 1-vortex moduli 
space $\M$ as a function of $\tau$, and show that the flow of the 
area with $\tau$ coincides with what occurs in Ricci flow. We
argue that for small $\tau$, the metric itself on $\M$ evolves from 
the metric on $\Sigma$ by Ricci flow. As is well-known \cite{Ma}, Ricci flow 
tends to smooth out the metric. For larger values of $\tau$ we cannot 
compute the metric on $\M$ in general. However, for the case of a 
vortex on a slightly deformed sphere of area $4\pi$, we calculate 
the asymptotic metric on $\M$ for $\tau$ approaching $1$.
The result is a round collapsing sphere, 
agreeing with what occurs in Ricci flow. For one vortex on a torus of
area $4\pi$ we show that no matter what the starting metric on the
torus, when $\tau$ reaches $1$, $\M$ is a flat torus. This is similar
to the result of Ricci flow, except that Ricci flow takes an infinite time to
produce a flat torus. These results suggest that in some precise way,
the geometric flow on the moduli space of one vortex is related to 
Ricci flow, but they are not conclusive. It is an open problem to 
obtain a general differential equation for the geometrical flow of 
the moduli space metric, and see whether or not it is equivalent 
to Ricci flow. 

We believe our results are of interest from the point of view of
connecting Ricci flow to the issue of the motion of finite-sized
objects in gravity \cite{Dix}. A basic axiom in gravity 
is that pointlike particles move along geodesics of space-time. 
Our vortices, being solitons, behave like particles, 
and when they are vanishingly small, and moving slowly, 
they move along geodesics of $\Sigma$. However, as $\tau$ increases 
from zero, the vortex size increases, and vortex motion is along 
a geodesic in the moduli space $\M$, which for one
vortex is still a path in $\Sigma$, but with a metric 
modified by Ricci flow. This gives a physical interpretation of
two-dimensional Ricci flow, at least for small times. It generates 
effective metrics for particles of small, finite size, whose geodesics 
are the approximate trajectories of these particles. This is a more
classical physical interpretation of Ricci flow than the known
interpretation involving the renormalisation group flow of sigma
models \cite{Frie}. For a recent discussion of possible physical
interpretations of geometrical flows, see ref.\cite{Kholo}.

\section{Vortex on a Surface}
\news

Let $\Sigma$ be a compact Riemann surface of genus $g$ with local
complex coordinate $z = x + iy$. We assume that $\Sigma$ has a
metric compatible with the complex structure 
\be
ds^2 = \Omega(x,y) (dx^2 + dy^2) = \Omega(x,y) dz d{\bar
  z} \,.
\ee
$\Omega$ is called the conformal factor.
The fields of the abelian Higgs model on $\Sigma$ are locally 
a complex scalar field $\phi$ and an abelian gauge potential whose components
$(a_x, a_y)$ combine naturally into a 1-form $a = a_x dx + a_y dy$. 
Globally, these are a section and $U(1)$ connection of a 
line bundle over $\Sigma$. The magnetic field is $B = \pr_x a_y -
\pr_y a_x$ and the first Chern number of the bundle is
\be
c_1 = \frac{1}{2\pi}\int_{\Sigma} B \, d^2x \,,
\ee
which is an integer that can be identified as both the number of
magnetic flux quanta and the net number of vortices on $\Sigma$.

This abelian Higgs model extends to a dynamical field theory on the
space-time $\Sigma \times \R$, with its product metric. We shall
consider the case where the scale parameters of the model are at 
``critical coupling''. For a discussion of the Lagrangian at critical 
coupling, including its kinetic and potential energy terms, see 
ref. \cite{ManSut}. Static solutions which minimize the potential energy
satisfy the first order Bogomolny equations \cite{Bog}
\bea
D_x\phi + iD_y\phi &=& 0 \,, \\
\frac{1}{\Omega} B - \half\left(\frac{1}{\tau} - |\phi|^2\right) &=& 0 \,,
\eea
where $D_i\phi = \pr_i\phi -ia_i\phi$, and $\tau$ is the positive 
(constant) Bradlow parameter.  

The pair of Bogomolny equations can be reduced to a single gauge
invariant equation as follows \cite{JafTau,Taubes}. With $z=x+iy$ and 
${\bar z} = x-iy$ we have $\pr_z = \half(\pr_x - i\pr_y)$ and 
$\pr_{\bar z} = \half(\pr_x + i\pr_y)$. Let $a_z = 
\half(a_x - ia_y)$ and $a_{\bar z} = \half(a_x + ia_y)$. The first
Bogomolny equation becomes $\pr_{\bar z}\phi - ia_{\bar z}\phi
= 0$, whose solution is
\be
a_{\bar z} = -i\pr_{\bar z}(\log \phi) \,.
\ee
$a_z$ is the complex conjugate of this. Now set
\be
\phi = e^{\half h + i\chi}
\ee
where $h$ is the gauge invariant and globally well-defined quantity
$\log|\phi^2|$. Then $B = -2i(\pr_z a_{\bar z} - \pr_{\bar z} a_z) 
= -2\pr_z\pr_{\bar z} h$. Substituting into the second Bogomolny
equation we obtain, for one vortex, Taubes' equation in the form
\be
\Delta h + \frac{1}{\tau} -
e^h = \frac{4\pi}{\Omega} \delta^{(2)}({\bf x} - {\bf X}) \,.
\label{Taubes}
\ee
$\Delta h \equiv \frac{4}{\Omega} \pr_z\pr_{\bar z} h$ is the covariant 
Laplacian of $h$ on $\Sigma$. The delta function source arises 
from the logarithmic singularity of $h$ at the vortex centre 
${\bf X}$, where $\phi$ vanishes and the magnetic flux density has its 
maximal value, $\frac{1}{2\tau}$. We denote by $Z$ the complex 
coordinate of the vortex centre, $Z= X +iY$. 

The key constraint on the existence of vortices arises by integrating
the second Bogomolny equation over $\Sigma$ (with the geometrical
measure $\Omega \, d^2x$), or equivalently, by integrating (\ref{Taubes}) 
(taking care over the logarithmic singularity). One finds, for one vortex,
\be
\frac{A}{2\tau} - \half\int_{\Sigma} |\phi^2| \Omega \, d^2x = 2\pi \,,
\label{IntBogo}
\ee
and since $|\phi^2|$ and $\Omega$ are non-negative, there is the
Bradlow inequality $\tau \le \frac{A}{4\pi}$. When $\tau$ approaches 
$\frac{A}{4\pi}$, the Higgs field becomes 
small everywhere and the magnetic flux density $\frac{1}{\Omega}B$ is 
almost uniform, that is, $B$ is approximately equal 
to $\frac{2\pi\Omega}{A}$. The vortex dissolves in the limit. 
The Bogomolny equations have solutions with uniform magnetic flux
density for $\tau = \frac{A}{4\pi}$, but $\phi$ vanishes 
identically, so the solutions are not vortices. Therefore, 1-vortex 
solutions of the Bogomolny equations exist only in the range
\be
0 < \tau < \frac{A}{4\pi} \,.
\label{Bradineq}
\ee
(Similarly, $N$-vortex solutions exist only for $0 < 
\tau < \frac{A}{4\pi N}$.)  
It is a result of Bradlow \cite{Brad} and Garc\'{\i}a-Prada \cite{Gar} that
for $\tau$ in this range, there is a
unique vortex solution for each choice of $Z$ on $\Sigma$.

It follows from the Lagrangian and Bogomolny argument (see
\cite{ManSut}) that the energy of one vortex is $E = \frac{\pi}{\tau}$. This 
can also be interpreted as the vortex mass. The size (that is, area) of a
vortex is of order $4\pi\tau$. This estimate comes from treating
the magnetic flux density as approximately $\frac{1}{2\tau}$ in the vortex core
and zero outside (really, it decays exponentially), and recalling that 
the total flux is $2\pi$.
The Bradlow inequality arises, intuitively, because a vortex of size
$4\pi\tau$ cannot be fitted into an area $A$ smaller than this. 

\section{The Vortex Moduli Space and its Metric}
\news

Provided $\tau$ satisfies the inequality (\ref{Bradineq}), there is a moduli
space of 1-vortex solutions $\M$, and as a manifold $\M = \Sigma$.
There is a natural metric on the moduli space. Mathematically, it is
the $L^2$ norm on fields tangent to the moduli space which are also
orthogonal to infinitesimal gauge transformations. Physically, 
it is derived from the expression for the kinetic energy of a 
vortex with slowly moving centre, $Z(t)$. Samols obtained a formula 
for the metric in terms of the local expansion of the field 
$h = \log|\phi|^2$ about the vortex centre \cite{Sam}. When 
the vortex is at $Z$, $h - 2\log|z-Z|$ is a real, regular function, so 
$h$ has an expansion
\be
h(z,{\bar z}) = 2\log|z-Z| + a + \half {\bar b} (z-Z) +  \half b ({\bar z} -
{\bar Z}) + \cdots \,,
\label{hexpan}
\ee
where $a$ and $b$ are functions of $Z$, ${\bar Z}$ and $\tau$. The kinetic
energy of the moving vortex is then
\be
T = \half \frac{\pi}{\tau} \left( \Omega + 2\tau \frac{\pr b}{\pr Z}
\right) {\dot Z}{\dot{\bar Z}}
\ee
with $\Omega$ evaluated at $Z$. Dropping the mass factor 
$\frac{\pi}{\tau}$ and the factor $\half$, we obtain the metric on $\M$,
\be
\left. ds^2\right|_{\M} = \left( \Omega + 2\tau \frac{\pr b}{\pr Z}
\right) dZ d{\bar Z} \,,
\label{Samols}
\ee
which, though defined using the local coordinate $Z$, can be shown 
to be globally consistent. This metric is a deformation of the 
original metric on $\Sigma$, the deformation being small when $\tau$
is small, as we will see below. The formula shows that globally, for 
all $\tau$, the metric on $\M$ is in the same conformal class as 
the metric on $\Sigma$.
 
It is not easy to determine $b$ in most situations, so, in general, 
the metric on $\M$ is not known. Some remarkable topological 
information about $b$ is however known, which we briefly review,
following \cite{ManNas}. Note first that $b$ is not invariant under changes
of coordinate. If, locally, we use a different complex coordinate
chart $z' = z'(z)$ on $\Sigma$, then $h$ has a similar expansion around
$Z'$ as (\ref{hexpan}),
\be
h(z',{\bar z}') = 2\log|z'-Z'| + a' + \half {\bar b}' (z'-Z') 
+  \half b' ({\bar z}' - {\bar Z}') + \cdots \,,
\ee
where, because of the logarithmic term, $a' = a - 
2\log|\frac{\pr z'}{\pr z}|$ and
\be
{\bar b}' = \left(\frac{\pr z'}{\pr z}\right)^{-1}{\bar b} -
\left(\frac{\pr z'}{\pr z}\right)^{-2}\frac{\pr^2 z'}{\pr z^2} \,,
\ee
with the quantities on the right hand side evaluated at $Z$. Let us 
define the 1-form $-{\bar b}(Z,{\bar Z}) \, dZ$, and its transformed 
version $-{\bar b}'(Z',{\bar Z}') \, dZ'$. Then
\be
-{\bar b}' \, dZ' = -{\bar b} \, dZ + 
\left(\frac{\pr Z'}{\pr Z}\right)^{-1}\frac{\pr^2 Z'}{\pr Z^2} \, dZ \,.
\ee
This is exactly the same transformation rule as for a connection
1-form on the holomorphic cotangent bundle of $\Sigma$, the canonical
bundle $K$, whose sections are locally $f(Z,{\bar Z}) \, dZ$ and where 
the transition rule is $f' = (\frac{\pr Z'}{\pr Z})^{-1} f$. Globally 
therefore, $-{\bar b} \, dZ$ is a connection 1-form on $K$,
varying with $\tau$. The Chern form of this connection is
\be
C_1(K) = \frac{i}{2\pi} d(-{\bar b} \, dZ) = 
\frac{i}{2\pi} \left(\frac{\pr {\bar b}}{\pr {\bar Z}}\right) \, dZ
\wedge d{\bar Z} \,.
\ee
For $\Sigma$ compact and of genus $g$, the integral of $C_1$ 
over $\Sigma$ is the Chern number, $c_1(K) = 2(g-1)$ \cite{GriHar}.

The K\"ahler 2-form associated to the metric (\ref{Samols}) is
\be
\omega = \frac{i}{2} \left(\Omega + 
2\tau\frac{\pr {\bar b}}{\pr {\bar Z}}\right) \, dZ
\wedge d{\bar Z} \,,
\ee
where we have used the reality property 
\be
\frac{\pr b}{\pr Z} = \frac{\pr {\bar b}}{\pr {\bar Z}}
\ee 
proved in \cite{ManSut}. Integrating
$\omega$ over $\Sigma$ gives the area of the 1-vortex moduli space $\M$,
\be
\left. A \right|_{\M} = A + 2\pi\tau c_1(K) = A + 4\pi\tau (g-1) \,,
\label{area}
\ee
where $A$ is the area of $\Sigma$. For all $g$, the range of the 
Bradlow parameter $\tau$ is as given in eq.(\ref{Bradineq}). The 
range of $\left. A\right|_{\M}$ is therefore from
$A$ to $gA$. When $g=0$, the area of $\M$ vanishes as $\tau \to
\frac{A}{4\pi}$ whereas for $g=1$ the area remains constant.

For a vortex on a sphere, plane or hyperbolic plane, in each
case with its standard metric of constant curvature, the metric on
$\M$ can be computed explicitly, and is \cite{Man2,Sam}
\be
\left. ds^2 \right|_{\M} 
= \left( 1 - \frac{\tau R_0}{2} \right)\Omega \, dZ d{\bar Z} \,,
\label{constcurv}
\ee
a rescaled version of the metric on the underlying surface $\Sigma$.
Here $R_0$ is the Ricci scalar curvature, which for a sphere of
radius $r$ is $\frac{2}{r^2}$. The result for the plane, with $R_0=0$, 
extends to any flat torus, but it is not known if the result for
the hyperbolic plane extends to a compact surface of constant negative
curvature. The metric (\ref{constcurv}) has been obtained using 
a symmetry argument to
find $b$, rather than solving the Taubes equation for $h$, but in the 
case when $R_0=-\frac{1}{\tau}$, Taubes' equation reduces to Liouville's 
equation, and the vortex solution and $b$ have a simple algebraic
form, leading explicitly to $\left. ds^2 \right|_{\M} 
= \frac{3}{2}\Omega \, dZ d{\bar Z}$ \cite{Strac}. 

\section{Vortex Moduli space for small $\tau$, and Ricci flow}
\news

Consider, as before, one vortex on a general surface $\Sigma$ with
smooth metric $ds^2 = \Omega \, dz d{\bar z}$. We give a simple argument which 
determines the metric on the moduli space $\M$ when $\tau$
is small. In this regime, the vortex is small, since its magnetic 
flux is concentrated in a region of area $4\pi\tau$, and outside 
this region the Higgs field is everywhere close to its
constant vacuum value, $|\phi|^2 = \frac{1}{\tau}$. The vortex
therefore only detects the local intrinsic curvature of $\Sigma$ at
the vortex centre $Z$, and it will move with kinetic energy equal to 
what the kinetic energy
would be on a surface of constant curvature. Using eq.(\ref{constcurv}) we
deduce that the metric on $\M$ is
\be
\left. ds^2 \right|_{\M} 
= \left( 1 - \frac{\tau R}{2} \right)\Omega \, dZ d{\bar Z} \,.
\label{smalltau}
\ee
Here, $R$ is the (non-uniform) Ricci scalar curvature evaluated at
$Z$, for which the formula is $R = -\frac{4}{\Omega}\pr_Z \pr_{\bar Z}
(\log \Omega)$. In the limit $\tau \to 0$, the metric on $\M$ smoothly
becomes the original metric on $\Sigma$, so geodesics on $\M$ become geodesics
on $\Sigma$, the result one expects for pointlike particles. 

We do not claim any rigour for this result, and corrections to the
metric of order $\tau^2$ are expected. A more careful asymptotic
analysis of the vortex solutions on $\Sigma$ as $\tau \to 0$ would be 
needed to prove it. The result is compatible with the exact formula 
(\ref{area}) for the area of $\M$, in the case that $\Sigma$ is 
compact, because the Gauss--Bonnet formula,
\be
\frac{i}{2}\int_{\Sigma} R \Omega  \, dZ \wedge d{\bar Z}
= 8\pi(1-g) \,,
\ee
implies that the area of $\M$ with the approximate metric (\ref{smalltau}) 
is $A + 4\pi\tau(g-1)$.
  
Let us now write the exact metric on $\M$ as
\be
\left. ds^2 \right|_{\M} = \Omega(\tau) \, dZ d{\bar Z} \,,
\ee
where $\Omega(0) = \Omega$. The expression (\ref{smalltau}) can 
be interpreted as saying that for small $\tau$, $\Omega(\tau)$ 
evolves by Ricci flow. This is because the Ricci flow equation on the surface
$\Sigma$ (with complex coordinate $Z$, and ``time'' $t$) is \cite{Ma}
\be
\frac{\pr}{\pr t} \Omega(t) = - \half R(t) \Omega(t) \,,
\label{Ricci}
\ee
where $R(t)$ is the Ricci scalar curvature of the metric with conformal 
factor $\Omega(t)$, and the dependence of $\Omega$ and $R$ on $Z$ and
${\bar Z}$ is implied. The initial condition is also $\Omega(0) =
\Omega$. The short-time solution of (\ref{Ricci}) is
\be
\Omega(t) = \left(1 - \frac{tR(0)}{2}\right)\Omega \,,
\ee
and this agrees with (\ref{smalltau}) if we identify $\tau$ with the 
time $t$ of the Ricci flow.

The area of the metric on $\Sigma$ under Ricci flow is 
$A + 4\pi t(g-1)$, which is derived from (\ref{Ricci}) by integrating 
over $\Sigma$ and using the Gauss--Bonnet formula. This agrees with the 
area of $\M$ for all $\tau$, if $\tau$ and $t$ are identified.
It would be very interesting if $\Omega(\tau)$ were precisely
the solution $\Omega(t)$ of the Ricci flow, with initial condition 
$\Omega$. To investigate this further, we shall consider the other
limit, where $\tau \to \frac{A}{4\pi}$ and the magnetic flux of the 
vortex is spread almost uniformly over the surface $\Sigma$. All our results 
scale in rather a simple way if the area $A$ and Bradlow parameter 
$\tau$ are multiplied by the same factor. So let us from now on fix $A=4\pi$, 
which implies that the range of $\tau$ is $0 < \tau < 1$.

\section{Linearized Ricci Flow on a Sphere}
\news

We need to review some results for the large-time asymptotics of
two-dimensional Ricci flow. In this section we consider $\Sigma$ 
a compact surface of genus zero, that is, $\Sigma$ is conformally a 2-sphere. 
The metric on $\Sigma$ can be written as 
\be
ds^2 = e^u \Omega_0 \, dz d{\bar z} \,,
\ee
where $z$ is a stereographic coordinate and $\Omega_0 
= 4/(1 + z {\bar z})^2$ defines the round metric on a sphere of unit
radius. $e^{u(z,{\bar z})}$ is the globally-defined conformal factor 
relative to the round sphere, so it should have a finite limit as 
$z \to \infty$. For the sphere to have area $4\pi$ we require
\be
\frac{i}{2} \int e^{u}\Omega_0 \, dz \wedge d{\bar z} = 4\pi \,.
\ee
Let us now assume that $u$ is small, and consider the Ricci flow of
this metric. We first find the normalised Ricci flow, with time
variable $\tilde t$, and then rescale the metric and time to obtain 
the true Ricci flow.

On a surface of genus zero and area $4\pi$ the averaged Ricci scalar
curvature is $2$. The normalised Ricci flow equation is therefore
\be
\frac{\pr \widetilde\Omega}{\pr {\tilde t}} = 
-\half ({\widetilde R} - 2) \widetilde\Omega \,,
\label{normRic}
\ee
where $\widetilde R$ is the Ricci curvature for the metric
$\widetilde\Omega \, dzd{\bar z}$. The round sphere with conformal 
factor $\Omega_0$ is a fixed point, so let us consider the 
linearized equation for metrics close by. Set
\be
\widetilde\Omega = e^u \Omega_0 \,.
\ee
The (exact) Ricci curvature is
\be
{\widetilde R} = e^{-u}(-\Delta_0 u + R_0) \,,
\ee
where $R_0=2$ is the scalar curvature for $\Omega_0$ 
and $\Delta_0$ is the covariant Laplacian operator in the
$\Omega_0$ background. To linear order in 
$u$, ${\widetilde R} = 2 - 2u - \Delta_0 u$, so the normalised Ricci 
flow equation is
\be
\frac{\pr u}{\pr {\tilde t}} = u + \half\Delta_0 u \,.
\ee
$u$ can be expanded in spherical harmonics, and for a harmonic $Y_j$ with
angular momentum $j$, $\Delta_0$ has eigenvalue
$-j(j+1)$. The coefficient $c_j$ of such a harmonic therefore satisfies
\be
\frac{dc_j}{d{\tilde t}} = \left(1 - \half j(j+1)\right)c_j \,,
\ee
so it generally decays exponentially with increasing ${\tilde t}$. 
The coefficient $c_0$
must vanish because of our assumption that the area of the deformed
sphere is $4\pi$. The coefficient $c_1$ can also be chosen to vanish,
as a non-zero value just corresponds to a coordinate
reparametrization of the sphere (since the curvature remains constant).
The non-trivial coefficient that decays most slowly is $c_2$, so 
generic deformed spheres approach the round sphere at late times 
with a deformation that is a $j=2$ spherical harmonic $Y_2$ (some 
real, linear combination of the five standard harmonics 
$Y_{2m}: \, -2 \le m \le 2$). If the initial 
metric is a round sphere with just a small deformation by a $j=2$ 
harmonic, $\widetilde\Omega(0) = (1 + Y_2)\Omega_0$, then the 
normalised Ricci flow gives
\be
\widetilde\Omega({\tilde t}) = (1 + e^{-2{\tilde t}}Y_2)\Omega_0 \,.
\label{linnormRic}
\ee

In two dimensions, Ricci flow is related to the
normalised Ricci flow by a simple scaling of the metric and a
reparametrisation of time, as follows. Suppose solutions of 
eqs.(\ref{Ricci}) and (\ref{normRic}) are related by
\be
\Omega = \mu \widetilde\Omega \,,
\ee
where $\mu$ depends only on time. Then $R = \frac{1}{\mu}{\widetilde R}$. 
Substituting in (\ref{Ricci}), one obtains
\be
\frac{\pr \widetilde\Omega}{\pr t} = -\frac{1}{\mu}\frac{d\mu}{dt}
\widetilde\Omega - \frac{1}{2\mu}{\widetilde R}\widetilde\Omega \,,
\ee
and hence
\be
\frac{\pr \widetilde\Omega}{\pr {\tilde t}} =
-\frac{1}{\mu}\frac{d\mu}{d{\tilde t}}\widetilde\Omega 
- \frac{1}{2\mu}\frac{dt}{d{\tilde t}}{\widetilde R}\widetilde\Omega \,.
\ee
This agrees with eq.(\ref{normRic}) provided 
\be
\frac{dt}{d{\tilde t}} = \mu \quad \quad {\rm and} \quad \quad 
-\frac{1}{\mu}\frac{d\mu}{d{\tilde t}} = 1 \,.
\ee 
The joint solution, assuming as initial condition that 
$\widetilde\Omega = \Omega$ at, respectively, ${\tilde t}=0$ and $t=0$, is 
\be 
\mu = e^{-{\tilde t}} \,,  \quad  t = 1 - e^{-{\tilde t}} \,.
\ee
As ${\tilde t}$ runs from 0 to $\infty$, $t$ runs from 0 to 1.
Applying this to the linearized solution of the normalised Ricci flow
(\ref{linnormRic}), we obtain the corresponding solution of the true Ricci flow
\be
\Omega(t) = (1-t)(1 + (1-t)^2 Y_2)\Omega_0 \,.
\ee
As $t$ approaches 1, the area of the sphere approaches zero linearly
with $t$, and the deformation relative to a round sphere approaches
zero quadratically.

We shall next study the metric on moduli space for one vortex on a
slightly deformed sphere, with deformation a $j=2$ harmonic $Y_2$. If the
geometrical flow of the moduli space metric were the same as Ricci
flow, then the conformal factor on $\M$ would be
\be
\Omega(\tau) = (1-\tau)(1 + (1-\tau)^2 Y_2)\Omega_0 \,.
\label{Ricdef}
\ee
We shall show that as the Bradlow parameter $\tau$ approaches 1, the area of
moduli space behaves as in (\ref{Ricdef}) and that the deformation of the
sphere does decay to zero at least linearly. But our calculation is
not refined enough to show that the deformation is proportional to
$(1-\tau)^2$. 

\section{Vortex on a deformed sphere}
\news

Explicit calculation of the metric on the 1-vortex moduli space $\M$
is generally hard, since the vortex solution is not explicitly known
even for a vortex on a plane. For a vortex on a round sphere the 
metric on $\M$ is known, but for a vortex on a deformed sphere it 
is not. Here we shall assume the deformation is by the simplest 
$j=2$ harmonic and small, and we shall work to linear
order in the deformation. We still cannot calculate the metric on $\M$
for the whole range of the Bradlow parameter $\tau$, so we 
concentrate on the case where $\tau$ is close to 1, complementing 
our earlier results where it was close to 0.

To explain our strategy, we first rederive the metric on moduli space 
for the case of a vortex on the round unit sphere with $\tau$ close to
$1$ \cite{Man2,BatMan}. It will be convenient to switch frequently 
between polar coordinates $\theta, \varphi$ and the stereographic 
coordinate $z= \tan\half\theta \, e^{i\varphi}$. We start with Taubes' equation
\be
\Delta_0 h + \frac{1}{\tau} - e^h =
\frac{4\pi}{\Omega_0} \delta^{(2)}({\bf x} - {\bf X})
\ee
where $\Delta_0 = \frac{4}{\Omega_0}\pr_z\pr_{\bar z} 
= (1 + z{\bar z})^2 \pr_z\pr_{\bar z}$ is the covariant Laplacian on
the unit sphere, and we set $\frac{1}{\tau} = 1 + \e$. For small 
$\e$, $e^h$ is small everywhere, of order $\e$, and vanishes at the vortex
centre. A consistent expansion is to set
\be
h = h_0 + \e h_1 + \e^2 h_2 + \cdots
\ee
where $h_0$ has a summand $\log \e$. $h_0$ satisfies 
\be
\Delta_0 h_0 + 1 = \frac{4\pi}{\Omega_0}
\delta^{(2)}({\bf x} - {\bf X} ) \,.
\ee 
For a vortex at the origin, ${\bf X} = {\bf 0}$,
the solution is
\be
h_0 = \log\frac{z{\bar z}}{1 + z{\bar z}} + \log C_0\e \,,
\ee
where the constant $C_0$ is still to be determined. Expanding out Taubes'
equation up to terms of order $\e$ we find the inhomogeneous equation 
for $h_1$,
\be
-\Delta_0 h_1 = 1 - C_0\frac{z{\bar z}}{1 + z{\bar z}} \,.
\label{h1eq}
\ee
In spherical polars, the operator on the left hand side is 
(minus) the standard Laplacian on a unit 2-sphere, and
\be
\frac{z{\bar z}}{1 + z{\bar z}} = \half(1 - \cos\theta) \,.
\ee 
The right hand side of (\ref{h1eq}) is required to have an expansion 
in spherical harmonics with no constant term, and this fixes $C_0=2$. 
The right hand side is then $\cos \theta$, a $j=1$ spherical
harmonic, so $h_1 = \half \cos \theta + C_1$. Expanding to next order 
in $\e$ we get an equation for $h_2$ which imposes a consistency 
condition on $C_1$. Although we do not actually need these, we find $C_1
= \frac{1}{6}$ and $h_2$ is a linear combination of the
Legendre polynomials $P_2(\cos\theta)$ and $P_1(\cos\theta)$
and an undetermined constant $C_2$. Proceeding further one could
construct a formal series expansion for $h$ in powers of $\e$, where 
each term is a finite polynomial in $\cos\theta$ or equivalently a 
finite sum of Legendre polynomials $P_m(\cos\theta)$. Converting back 
to a function of $z$ and ${\bar z}$, our solution for $h$, up 
to order $\e$, is
\be
h= \log\frac{z{\bar z}}{1 + z{\bar z}}  + \log2\e 
+ \e\left(\frac{2}{3} - \frac{z{\bar z}}{1 + z{\bar z}}\right) 
+ \cdots \,.
\ee

Spherical symmetry means that it is easy to convert this solution into
the solution for a vortex located at a general point ${\bf X}$ on the sphere. 
Let $Z$ be the stereographic coordinate corresponding to ${\bf X}$. We
observe that $\frac{4z{\bar z}}{1 + z{\bar z}}$ is the square of the chordal
distance from $z$ to 0 on the unit sphere, and the rotated version of 
this function is
\be
\frac{4(z-Z)({\bar z} - {\bar Z})}{(1 + z{\bar z})(1 + Z{\bar Z})} \,,
\label{chordal}
\ee
the square of the chordal distance between $z$ and $Z$. Therefore, 
for the vortex centred at $Z$,
\be
h = \log \frac{(z-Z)({\bar z} - {\bar Z})}{(1 + z{\bar z})(1 + Z{\bar Z})}
 + \log 2\e + \e \left( \frac{2}{3} - \frac{(z-Z)({\bar z} - {\bar
     Z})}{(1 + z{\bar z})(1 + Z{\bar Z})} \right) + \cdots \,.
\ee
Using this, we can calculate the metric on moduli space. We need the partial
derivative with respect to ${\bar z}$ of $h - 2\log|z-Z|$
evaluated at $Z$. Only the term $-\log(1 + z{\bar z})$ contributes to
this, so
\bea
b(Z,{\bar Z}) &=& 
\left. 2 \frac{\pr}{\pr {\bar z}}(-\log(1 + z{\bar z}))\right|_{z=Z} 
\nonumber \\
     &=& - \frac{2Z}{1 + Z{\bar Z}} \,,
\eea
and hence
\be
\frac{\pr b}{\pr Z} = -\frac{2}{(1 + Z{\bar Z})^2} \,.
\ee
The Samols metric (\ref{Samols}) is therefore, to order $\e$,
\be
\left. ds^2 \right|_{\M} = \left(\frac{4}{(1 + Z{\bar Z})^2} 
- (1-\e)\frac{4}{(1 + Z{\bar Z})^2}\right)dZ d{\bar Z}
= \frac{4\e \, dZ d{\bar Z}}{(1 + Z{\bar Z})^2} \,.
\ee
This is just a scaled version of the metric on the round sphere where 
the vortex resides, and the area goes to zero as $\e \to 0$. The
result agrees with (\ref{constcurv}), with $R_0=2$ and 
$\frac{1}{\tau} = 1 + \e$.

Let us now consider the vortex on the deformed sphere with conformal
factor
\be
\Omega = (1 + \alpha P_2(\cos\theta))\Omega_0 \,,
\ee
where $\alpha$ is small. This ellipsoidal deformation by the $j=2$
harmonic $P_2(\cos\theta)$ is axially symmetric in $\varphi$. We
set $\frac{1}{\tau} = 1 + \e$ as before. We work to linear order in
$\alpha$, and in principle to arbitrary order in $\e$ (i.e. we consider a
vortex of arbitrary size on the slightly deformed sphere). In practice,
we shall just calculate up to linear order in $\e$ to establish the
metric on moduli space close to $\tau = 1$. 

Note that the chordal distance (\ref{chordal}) can be written as
\be
2(1-{\bf n}_z \cdot {\bf n}_Z) \,,
\label{chordal1}
\ee
where ${\bf n}_z = (\sin\theta \cos\varphi, \sin\theta \sin\varphi,
\cos\theta)$ is the unit Cartesian vector corresponding to $z$, and
${\bf n}_Z$ similarly corresponds to $Z$. If $Z$ is real, we can write
${\bf n}_Z = (\sin\lambda , 0, \cos\lambda)$, and then  
\be
2(1-{\bf n}_z \cdot {\bf n}_Z) = 2(1 - \sin\lambda
\sin\theta \cos\varphi - \cos\lambda\cos\theta) \,.
\ee 
As a function of $\theta$ and $\varphi$ this is a linear combination 
of $j=0$ and $j=1$ harmonics. The deformation of the sphere by a
$j=2$ harmonic will generate products of $j=1$ and $j=2$
harmonics. Our calculation of $h$ can proceed because these products 
are themselves finite sums of harmonics with $j=1,2,3$, by the 
usual Clebsch--Gordon rules.

The Taubes equation (\ref{Taubes}) on the deformed sphere can be written as
\be
\Delta_0 h + (1 + \e - e^h)(1 + \alpha P_2(\cos\theta)) =
\frac{4\pi}{\Omega_0} \delta^{(2)}({\bf x - \bf X}) \,.
\label{Taubesdef}
\ee
We will solve this as before, inverting $\Delta_0$, the Laplacian on the round
sphere, on a basis of spherical harmonics. We start again with the case
${\bf X} = {\bf 0}$, and set
$h = h_0 + \e h_1 + \cdots$ where $h_0$ has a $\log\e$ term ensuring
that $e^h$ is of order $\e$. $h_0$ is the solution of eq.(\ref{Taubesdef}) 
with the terms $\e - e^h$ dropped but the term proportional to $\alpha$
retained. The solution, this time in trigonometric form, is 
\be
h_0 = \log\half(1 - \cos\theta) + \log C_0\e 
+ \frac{\alpha}{6} P_2(\cos\theta) \,,
\label{h0soln}
\ee
with $C_0$ to be determined. The equation for $h_1$ is obtained from
the terms of order $\e$ in (\ref{Taubesdef}), noting that $e^h$ can be
replaced by $e^{h_0}$ at this order. There results the inhomogeneous equation
\be
-\Delta_0 h_1 = \left(1 - \frac{C_0}{2}\left(1 - \cos\theta\right)
\left(1 + \frac{\alpha}{6}P_2(\cos\theta)\right)\right)
\left(1 + \alpha P_2(\cos\theta)\right) \,,
\label{h1def}
\ee  
from which terms quadratic in $\alpha$ are dropped. 
The right hand side is expressible as a linear combination of 
$P_m(\cos\theta)$ with $m=1,2,3$, and no constant term, provided $C_0
= 2$ as before. Eq.(\ref{h1def}) then simplifies to
\be
-\Delta_0 h_1 = \left(1 + \frac{7\alpha}{15}\right)P_1(\cos\theta) 
- \frac{\alpha}{6}P_2(\cos\theta) 
+ \frac{7\alpha}{10}P_3(\cos\theta) \,,
\ee
whose solution is
\be
h_1 = \left(\half + \frac{7\alpha}{30}\right)P_1(\cos\theta)
-\frac{\alpha}{36}P_2(\cos\theta) + \frac{7\alpha}{120} P_3(\cos\theta)
+ C_1 \,.
\ee
The constant $C_1$ can be determined using the equation for $h_2$.

Now we can tackle the general case, with ${\bf X}$ arbitrary. The equation
for $h_0$ differs from the round sphere case by a term of order $\alpha$
and the solution is
\be
h_0 = \log \frac{(z-Z)({\bar z} - {\bar Z})}{(1 + z{\bar z})(1 +
Z{\bar Z})} + \log C_0\e + \frac{\alpha}{6}P_2(\cos\theta) \,.
\ee
We recall that 
\be
\frac{(z-Z)({\bar z} - {\bar Z})}{(1 + z{\bar z})(1 + Z{\bar Z})} 
=  \half (1 - \sin\lambda \sin\theta \cos\varphi 
- \cos\lambda\cos\theta) \,,
\label{polarconv}
\ee
and hence the equation for $h_1$ is 
\be
-\Delta_0 h_1 = 1 + \alpha P_2(\cos\theta)
- \frac{C_0}{2}(1 - \sin\lambda\sin\theta \cos\varphi - \cos\lambda\cos\theta)
\left(1 + \frac{7\alpha}{6} P_2(\cos\theta)\right) \,.
\ee
Multiplying this out, we find yet again that $C_0 = 2$, and then 
we can express the right hand side in terms of spherical harmonics 
with $j=1,2,3$. Inverting the Laplacian we find
\bea
h_1 &=& \half\sin\lambda\sin\theta\cos\varphi +
\half\cos\lambda\cos\theta - \frac{\alpha}{36} P_2(\cos\theta) 
\nonumber \\
&+& \frac{7\alpha}{6}\cos\lambda\left(\frac{1}{20}P_3(\cos\theta) + 
\frac{1}{5}P_1(\cos\theta)\right)
\nonumber \\ 
&+& \frac{7\alpha}{6}\sin\lambda\left(\frac{1}{40}Y_3 -
\frac{1}{10}\sin\theta \cos\varphi\right) + C_1
\eea
where $Y_3 = (5\cos^2\theta\sin\theta - \sin\theta)\cos\varphi$ 
is a $j=3$ harmonic. This can be reorganised in the form
\be
h_1 = \half\Lambda - \frac{\alpha}{36} P_2(\cos\theta) + 
\frac{7\alpha}{24}\left(\cos\lambda\cos\theta - \half\Lambda\sin^2\theta\right)
 + C_1
\label{h1soln}
\ee
where $\Lambda = \sin\lambda\sin\theta \cos\varphi + \cos\lambda\cos\theta$.

Combining (\ref{h0soln}) and (\ref{h1soln}) gives $h= h_0 + \e h_1 + \dots$ to
order $\e$. Next we convert $h$ to a function of $z$ and ${\bar z}$,
using (\ref{polarconv}) and $\cos\theta = \frac{1 - z{\bar z}}{1 + z{\bar z}}$.
We then remove the logarithmically singular term, defining
${\tilde h} = h - 2\log|z - Z|$, and calculate the Samols coefficient
\be
b = \left. 2\frac{\pr}{\pr {\bar z}} {\tilde h} \right|_{z=Z} \,.
\ee
This simplifies to
\be
b = -\frac{2Z}{1 + Z{\bar Z}} - 2\alpha (1 + \e)
\frac{(1 - Z{\bar Z})Z}{(1 + Z{\bar Z})^3} \,,
\label{bdef}
\ee
where we have used $\cos\lambda = \frac{1 - Z{\bar Z}}{1 + Z{\bar Z}}$.  
The term proportional to $\alpha$ is new here. Expression (\ref{bdef})
has been obtained assuming $Z$ is real, but by axial symmetry it is
valid for all $Z$. Differentiating again, we find
\be
\frac{\pr b}{\pr Z} = - \frac{2}{(1 + Z{\bar Z})^2} 
- 2\alpha (1 + \e)\frac{1 - 4Z{\bar Z} + (Z{\bar Z})^2}
{(1 + Z{\bar Z})^4} \,,
\ee
where the final expression is equivalently
\be
\frac{1 - 4Z{\bar Z} + (Z{\bar Z})^2}
{(1 + Z{\bar Z})^4} = \frac{P_2(\cos\lambda)}{(1 + Z{\bar Z})^2} \,.
\ee
Finally, we calculate the metric on moduli space $\M$,
\be
\left. ds^2 \right|_{\M} 
= \left( \Omega + 2\tau\frac{\pr b}{\pr Z}\right)dZ d{\bar Z} \,,
\ee
using $\Omega = (1 + \alpha P_2(\cos\lambda))\Omega_0$ and the
approximation $\tau = 1 - \e$. The terms not proportional to $\e$
cancel, as do the terms involving $P_2(\cos\lambda)$, leaving
\be
\left. ds^2 \right|_{\M} = \frac{4\e \, dZ d{\bar Z}}{(1 + Z{\bar Z})^2}
\ee
with order $\e^2$ corrections which could be found by systematically
extending this calculation. As expected, the area of the moduli space 
is $\e$ times the area of $\Sigma$, and vanishes as $\e \to 0$. 
Remarkably, the deformation of the sphere also
vanishes from the moduli space at this order. These conclusions are 
in agreement with what occurs in Ricci flow. Further calculation might 
show a deformation relative to the round sphere of order $\e$,
although agreement with Ricci flow would require a deformation only 
at order $\e^2$.

\section{Vortex on a torus}
\news

On a surface of genus 1, a torus, Ricci flow and
normalised Ricci flow are the same. Given any starting
metric, Ricci flow takes it to a flat metric, while preserving the
conformal class and the area of the torus. From the linearized Ricci
flow equation, one sees that the approach to the flat metric
takes infinite time.

Consider now one vortex on a torus $\Sigma$ with local complex
coordinate $z=x+iy$. We simplify our analysis by choosing a square
torus, $\{(x,y): 0 \le x \le 1 \,, 0 \le y \le 1\}$ with opposite sides
identified. The metric is $ds^2 = \Omega \, dzd{\bar z}$ with
$\Omega$ a smooth function on the torus, and we assume 
\be
\int_0^1 \int_0^1 \Omega \, dxdy = 4\pi \,,
\ee
so the area of the torus is $4\pi$. The Higgs field and gauge
potential are a section and connection of a $U(1)$ bundle with Chern
number 1. We take the bundle to be defined by the transition function 
$g(x) = e^{-2\pi ix}$ relating the top edge $(y=1)$ to the bottom 
edge $(y=0)$, and by the trivial transition function relating
the right edge $(x=1)$ to the left edge $(x=0)$.

As always, the 1-vortex moduli space $\M$ is a copy of $\Sigma$, 
with the vortex centre $Z$ as coordinate, and metric
\be
\left. ds^2\right|_{\M} = \Omega(\tau) \, dZd{\bar Z}
= \left(\Omega + 2\tau\frac{\pr b}{\pr Z}\right)\, dZd{\bar Z} \,,
\label{Samolstor}
\ee
with notation as before. The cotangent bundle of the torus is trivial,
so $b$ is a well-defined, smooth function on $\M$. 

Our earlier discussion shows that for small $\tau$, $\Omega(\tau)$
evolves from the initial metric $\Omega$ according to the Ricci
flow. Moreover, it follows from (\ref{Samolstor}) that for all $\tau$, $\M$ is 
conformally a square torus and also, since the Chern number of the 
cotangent bundle vanishes, that the area of $\M$ is $4\pi$, the same
as the area of $\Sigma$. We shall show in this section that in the 
Bradlow limit $\tau \to 1$, $\Omega(\tau)$ becomes the flat metric. 
All this is similar to the Ricci flow, with one important difference. 
$\Omega(\tau)$ is defined only on the finite interval $0 < \tau < 1$. 
The time $t$ in the Ricci flow, and the Bradlow parameter $\tau$, 
cannot therefore be identified.

It is easiest to understand what happens in the strict Bradlow limit, 
$\tau = 1$. The moduli space does not collapse to a point here, as it 
did for a vortex on a sphere. The Bogomolny equations reduce to 
\be
B = \half \Omega \,,
\label{mag}
\ee
and $\phi$ vanishes everywhere. However, the connection is not
completely determined by (\ref{mag}). Given a 1-form gauge potential 
$a^{(0)}$, the general one is $a = a^{(0)} + \alpha$ where 
$\alpha$ has zero magnetic field, i.e. 
$\alpha$ is a flat connection on $\Sigma$. By a gauge choice we can assume
\be
\alpha = 2\pi(\mu dx + \nu dy)
\ee
where $\mu$ and $\nu$ are independent of $x$ and $y$. Furthermore, we
can restrict $\mu,\nu$ to the ranges $-\half \le \mu \le \half$, 
$-\half \le \nu \le \half$ (with endpoints identified), since a legitimate
gauge transformation on the torus is $g(x,y) = e^{2\pi i (mx +ny)}$
with $m,n$ integers, and this shifts $\mu$ by $m$ and  
$\nu$ by $n$. The moduli space $\M$ at $\tau = 1$ is therefore
the square torus $\left\{(\mu,\nu):  -\half \le \mu \le \half \,, 
-\half \le \nu \le \half \right\}$. The metric on $\M$ can be directly
obtained from the kinetic part of the Lagrangian for time-dependent
$\mu$ and $\nu$. This is
\be
T = \half \int_0^1 \int_0^1 (e_x^2 + e_y^2) \, dxdy \,,
\ee
where
\be
e_x = {\dot a}_x - \pr_x a_0 \,, \quad e_y = {\dot a}_y - \pr_y a_0
\ee
are the electric field components, with $a_0$ the time component of
the gauge potential. Note that factors of
$\Omega$ cancel out in $T$. In our gauge, it is consistent with Gauss'
law, $\pr_x e_x + \pr_y e_y = 0$, to set $a_0 = 0$, so $e_x = 2\pi
\dot\mu$ and $e_y = 2\pi \dot\nu$, and $T$ reduces to
\be
T = 2\pi^2 (\dot\mu^2 + \dot\nu^2) \,.
\ee
The metric on moduli space is therefore 
\be
\left. ds^2\right|_{\M} = 4\pi(d\mu^2 + d\nu^2) \,,
\label{flatmet}
\ee
where we have extracted the vortex mass $\pi$ in the limit $\tau = 1$ (the
energy in the magnetic field and Higgs field) and the factor $\half$. 
The metric is clearly flat, whatever $\Omega$ was initially, and the area of 
$\M$ is $4\pi$ as expected.

Before moving on, it is convenient to note here how the moduli of the
flat connection can be described in an alternative gauge. Suppose we
perform the gauge transformation $g(x,y)= e^{-2\pi i(\mu x + \nu y)}$. 
$\alpha$ now vanishes, but instead, the bundle transition functions
are $g(x)= e^{-2\pi i(x+\nu)}$ relating $y=1$ to $y=0$, and 
$g(y) = e^{-2\pi i\mu}$ relating $x=1$ to $x=0$. For time-dependent
$\mu$ and $\nu$, $a_0 = -2\pi(\dot\mu x + \dot\nu y)$, consistent with
$a_0$ on opposite edges of the square being related by the time
derivative of the transition functions, and the electric
field is as before.

It is fairly clear that the metric on $\M$ smoothly approaches the
limiting flat metric (\ref{flatmet}) as $\tau$ approaches $1$. This is 
because the magnetic field approaches the limiting value $\frac{\Omega}{2}$, 
and the Higgs field becomes vanishing small. However, what needs clarification
is how the time-varying moduli $\mu$ and $\nu$ are related to the
actual motion of the vortex, that is, the motion of its centre $Z$.
We shall now show that $Z$ is linearly related to $\mu$ and $\nu$ 
when $\tau$ is close to $1$, and that the
moduli space $\M$ has the canonical flat metric with $Z$ as
complex coordinate. We shall calculate this metric just at zeroth order
in $\e = \frac{1}{\tau} - 1$. Our method is rather close to that of
refs.\cite{GonRam,GonRam1}.  

At zeroth order, $B = \frac{\Omega}{2}$. We rewrite this as
\be
B = 2\pi + \frac{\tilde\Omega}{2}
\ee
where $\tilde\Omega$ integrates to zero. Then one
choice of gauge potential is
\be
a_x = -2\pi y - \half\pr_y{\tilde K} \,, \quad 
a_y = \half\pr_x{\tilde K} \,,
\label{gaugetor}
\ee
where $\nabla^2{\tilde K} = \tilde\Omega$. This Poisson equation has a
unique solution for $\tilde K$ on the torus, up to an irrelevant
additive constant. The gauge potential (\ref{gaugetor}) is consistent with
either of the gauge choices we introduced above, and we choose the
second of these, with the moduli $\mu$ and $\nu$ present in the 
transition functions. 

The moduli are physical, since they affect the holonomy of the 
gauge potential (\ref{gaugetor}) around cycles in the $x$- and $y$-directions. 
More importantly for us, they affect the Higgs field. Recall the 
first Bogomolny equation 
\be
D_{\bar z}\phi \equiv \half (\pr_x + i\pr_y - ia_x + a_y)\phi = 0 \,.
\label{Bogo1}
\ee
We need to solve this with the above gauge potential, and boundary
conditions
\bea
\phi(x,y=1) &=& \phi(x,y=0)e^{-2\pi i(x+\nu)} \,, 
\nonumber \\
\phi(x=1,y) &=& \phi(x=0,y)e^{-2\pi i\mu} \,.
\label{torBC}
\eea
We may use the integrating factor $K = \pi y^2 + \half {\tilde K}$,
and set $\phi = e^{-K} \psi$. Then (\ref{Bogo1}) reduces to 
$\pr_{\bar z} \psi = 0$, so $\psi(z)$ is holomorphic. The boundary 
conditions (\ref{torBC}) become
\bea
\psi(z+1) &=& \psi(z)e^{-2\pi i\mu} \,, 
\nonumber \\
\psi(z+i) &=& \psi(z)e^{-2\pi iz}e^{\pi}e^{-2\pi i\nu} \,.
\eea
These are exactly the defining equations for a theta-function with
characteristics on a square torus \cite{FarKra}, so
\be
\psi(z) = C \Theta_{-2\mu \brack 2\nu} (z; i) \,.
\ee
$C$ is a normalisation constant depending on $Z$ and ${\bar Z}$. Its
phase is arbitrary, but $|C|^2$ is determined by eq.(\ref{IntBogo}) 
and is of order $\e$. The theta-function has one zero in the
unit square, at
\be
Z = \frac{1-2\nu}{2} + \frac{1+2\mu}{2} \, i
\ee
and this is the vortex centre. So, $Z$ is linearly related to 
$\mu$ and $\nu$, and if $Z$ moves at constant velocity, there is 
a constant electric field orthogonal to the velocity.

Now we can use Samols' formula to find the metric on moduli
space. As usual, we need to expand $h = \log|\phi|^2$ around $Z$.
We set $z = Z + w$ and use the identity
\bea
&& \Theta_{-2\mu \brack 2\nu} \left(\frac{1-2\nu}{2} + \frac{1+2\mu}{2} i 
+ w ; i \right)
\nonumber \\
&& \quad \quad \quad \quad \quad 
= e^{2\pi i\left[-\half(1+2\mu)w - \frac{1}{8}(1 + 2\mu)^2 i -
    \frac{1}{4}(1+2\mu)\right]} \Theta_{1 \brack 1} (w ; i) \,,
\eea
where the final theta-function has the expansion around $w=0$, 
\be
\Theta_{1 \brack 1} (w ; i) = \alpha w(1 + \gamma w^2 + \cdots) \,.
\ee
Therefore,
\be
h = -2\pi y^2 - {\tilde K} + 2\log|C| + \frac{\pi}{2}(1 + 2\mu)^2 +
2\pi(1 + 2\mu) {\rm Im} \, w + 2\log|\alpha| + 2\log|w| + \cdots \,.
\ee
Since $y = \frac{1+2\mu}{2} + {\rm Im} \, w$, this expansion of $h$ simplifies
to the desired form
\be
h = 2\log|w| + a + \half{\bar b} w + \half b {\bar w} + \cdots
\ee
where
\be
b = \left. -2\pr_{\bar w}{\tilde K} \right|_{w=0}
= \left. -2\pr_{\bar z}{\tilde K} \right|_{z=Z} \,.
\ee
Therefore, using the Samols formula (\ref{Samolstor}),
\be
\left. ds^2\right|_{\M} = (\Omega - \tau \nabla^2{\tilde K})\,
dZd{\bar Z} \,,
\ee
with $\Omega$ and $\tilde K$ here regarded as functions of $Z$
and $\bar Z$. But recall that $\nabla^2{\tilde K} = \Omega - 4\pi$, 
and $\tau \simeq 1 - \e$, so to zeroth order in $\e$,
\be
\left. ds^2\right|_{\M} = 4\pi \, dZd{\bar Z} \,.
\ee

This result in the case that $\Sigma$ is a flat
torus has been obtained previously. What is novel is that in the
Bradlow limit, to zeroth order in $\e$, the contribution of $\tilde
K$, which accounts for the deformation of the metric on $\Sigma$, 
cancels out in the metric on the moduli space. Calculating the metric 
on $\M$ to order $\e$, for any non-trivial $\tilde K$, would be a 
tricky exercise in theta-functions, and we have not attempted it.

\section{Conclusion}
\news 

For one abelian Higgs vortex on a compact surface $\Sigma$
with arbitrary metric, we have studied the metric on the moduli space
$\M$. The metric on $\M$ is in the same conformal class as the metric
on $\Sigma$, and it has been shown to exhibit an interesting 
geometrical evolution as the Bradlow parameter $\tau$ varies. For 
small $\tau$, the metric evolves from the original metric on $\Sigma$ 
in the same way as in Ricci flow. Further similarities to Ricci flow 
have been established by studying a vortex on a slightly deformed 
sphere, and on a deformed square torus, as $\tau$ approaches the 
Bradlow limit (which is $\tau = 1$ if $\Sigma$ has area $4\pi$). 
Possibly there is a precise relation for all $\tau$ between 
the geometrical flow of the moduli 
space metric and Ricci flow, although the example of a vortex on a 
torus shows that there must be at least a difference in the time 
parametrization. It would be interesting to understand the metric on the
moduli space in the Bradlow limit in the case that the genus $g$
of $\Sigma$ is greater than 1, since this metric appears to depend only on
the conformal class of $\Sigma$. Nasir has studied the metric for
$N$-vortices with $N \ge g$ in the Bradlow limit, but did not obtain 
an explicit result in the 1-vortex case \cite{Nas}. 

Since vortex motion is approximately along a geodesic in moduli space,
our results suggest that the metrics generated by Ricci 
flow have a physical interpretation as effective metrics whose 
geodesics model the motion of particles of finite size. These metrics 
replace the starting metric whose geodesics model the motion of 
pointlike particles.

\section*{Acknowledgement}

I would like to thank Vladimir Matveev for a useful discussion.

\end{document}